\documentstyle[epsfig,12pt]{article}
\newcommand{\be}{\begin{eqnarray}}
\newcommand{\ee}{\end{eqnarray}}

\def\lsim{\mathrel{\rlap{\lower4pt\hbox{\hskip1pt$\sim$}}
    \raise1pt\hbox{$<$}}}               
\def\gsim{\mathrel{\rlap{\lower4pt\hbox{\hskip1pt$\sim$}}
    \raise1pt\hbox{$>$}}}               

\begin{document}

\rightline{\Large{Preprint RM3-TH/00-7}}

\vspace{2cm}

\begin{center}

\LARGE{Soft gluon effects in the extraction of\\ higher twists at large Bjorken x\footnote{\bf To appear in Physics Letters B.}}

\vspace{1cm}

\large{Silvano Simula}

\vspace{0.5cm}

\normalsize{Istituto Nazionale di Fisica Nucleare, Sezione Roma III\\ Via della Vasca Navale 84, I-00146 Roma, Italy}

\end{center}

\vspace{1cm}

\begin{abstract}

\noindent Existing data on the (unpolarised) transverse structure function of the proton are analyzed for large values of the Bjorken variable $x$. The leading-twist and a phenomenological higher-twist contributions are simultaneously determined from a power correction analysis of the Nachtmann moments for values of the squared four-momentum transfer between $\sim 1$ and $20 ~ (GeV/c)^2$. The results obtained adopting the next-to-leading order approximation and those including the effects of soft gluon resummation are compared. The sensitivity of the extraction of large-$x$ higher twists to high-order radiative corrections as well as to the value of $\alpha_s(M_Z^2)$ is illustrated.

\end{abstract}

\newpage

\pagestyle{plain}

\indent The experimental and theoretical investigation of the behaviour of parton distribution functions ($PDF$) for large values of the Bjorken variable $x$ as well as the extraction of higher twist terms in the Deep Inelastic Scattering ($DIS$) structure functions of the nucleon has recently received a renewed attention \cite{Bodek}-\cite{Alekhin}. The main goals are: ~ i) the determination of multiparton correlations in the nucleon \cite{Ji}-\cite{Alekhin}, and ~ ii) the investigation of ratio of $d$- to $u$-quark distributions in the vicinity of the threshold region $x \simeq 1$ \cite{Bodek,Ricco}. In this way one may test not only models of the structure of the nucleon (see, e.g., \cite{Isgur}), but also the predictions of perturbative $QCD$ at large values of $x$, which can be relevant for the experimental programs at high-energy facilities, like $CEBAF$, $HERA$ and future $LHC$.

\indent The aim of this letter is to address the issue of the relevance of high-order radiative corrections in the extraction of both the leading- and the higher-twist contributions from the (unpolarised) $DIS$ structure function of the proton for large values of the Bjorken variable $x$. To this end: ~ i) we perform power correction analyses of the Nachtmann moments of the transverse proton structure function, $M_n(Q^2)$, constructed in \cite{Ricco} starting from world data in $DIS$ kinematics as well as in the nucleon-resonance production regions for values of the squared four-momentum transfer $Q^2$ between $\sim 1$ and $20 ~ (GeV/c)^2$ \footnote{Note that existing $SLAC$ data up to $x \simeq 0.98$ as well as the elastic peak contribution are included in the construction of the Nachtmann moments performed in \cite{Ricco}. Moreover, the inclusion of data in the nucleon-resonance production region is a relevant point in view both of the extraction of multiparton correlations (target-dependent higher twists) and of parton-hadron duality arguments (see again  \cite{Ricco}).}$^,$\footnote{We want to remind that the Nachtmann moments of Ref. \cite{Ricco} have been obtained using available interpolation formulae for evaluating the nucleon structure functions at fixed values of $Q^2$. In this respect new systematic measurements are needed to confirm the $Q^2$-behaviour of the Nachtmann moments found in \cite{Ricco}.}; and ~ ii) we compare the results obtained adopting a fixed-order calculation with those including the effects of soft gluon resummation ($SGR$) \cite{Sterman,Catani_1}. The Nachtmann moments $M_n(Q^2)$ will be analyzed in terms of the following twist expansion (cf. \cite{Ricco})
 \be
       M_n(Q^2) = \mu_n(Q^2) + {Q_0^2 \over Q^2} ~ a_n^{(4)} \left[ 
       {\alpha_s(Q^2) \over \alpha_s(Q_0^2)} \right]^{\gamma_n^{(4)}} + 
       {Q_0^4 \over Q^4} ~ a_n^{(6)} \left[ {\alpha_s(Q^2) \over 
       \alpha_s(Q_0^2)} \right]^{\gamma_n^{(6)}}
       \label{eq:twists}
 \ee
where $\mu_n(Q^2)$ is the twist-2 contribution, while the $pQCD$ evolution of the twist-4 [twist-6] is accounted for by the term $(\alpha_s(Q^2) / \alpha_s(Q_0^2))^{\gamma_n^{(4)}}$ [$(\alpha_s(Q^2) / \alpha_s(Q_0^2))^{\gamma_n^{(6)}}$] with an effective anomalous dimension $\gamma_n^{(4)}$ [$\gamma_n^{(6)}$] (cf. also \cite{Ji}) and the parameter $a_n^{(4)}$ [$a_n^{(6)}$] represents the overall strength of the twist-4 [twist-6] term at the chosen scale $Q_0^2$. In what follows we will limit ourselves to moments with order $n \geq 4$, mainly because the second moment is only slightly sensitive to the large-$x$ region and it turns out to be almost totally dominated by the leading twist (cf. \cite{Ricco}). Moreover, the effects of soft gluon resummation for the second moment turn out to be quite small (see later on Fig. 1). Finally, since at large $x$ the evolution of the singlet quark and gluon operators decouple, we consider the non-singlet ($NS$) evolution for the full twist-2 term (cf., e.g., \cite{Ricco}).

\indent For the fixed-order calculation of the leading twist we consider the next-to-leading ($NLO$) order approximation, viz. (cf. \cite{BB79})
 \be
       \mu_n^{(NLO)}(Q^2) = A_n^{(NLO)} \left[ {\alpha_s(Q^2) \over 
       \alpha_s(Q_0^2)} \right]^{\gamma_n^{NS}} ~ \left[ 1 + {\alpha_s(Q^2) 
       \over 4 \pi} \left( 2 C_n^{(NLO)} + \gamma_n^{(1, NS)} - {\beta_1 
       \over \beta_0}  \gamma_n^{NS} \right) \right]
       \label{eq:NLO}
 \ee
where $\gamma_n^{NS}$ are the one-loop $NS$ anomalous dimensions, $\gamma_n^{(1, NS)}$ are the two-loop $NS$ anomalous dimensions, $\beta_0 = 11 - 2 N_f / 3$, $\beta_1 = 102 - 38 N_f / 3$ (with $N_f$ representing the number of active flavours) and $A_n^{(NLO)}$ is the n-th moment of $PDF$ at the scale $Q_0^2$. In Eq. (\ref{eq:NLO}) $C_n^{(NLO)}$ is the $NLO$ part of the quark coefficient function, that in the $\overline{MS}$ scheme reads as
 \be
      C_n^{(NLO)} = C_F \left\{ S_1(n) \left[ S_1(n) + {3 \over 2} - {1 
      \over n(n+1)} \right] - S_2(n) + {3 \over 2n} + {2 \over n+1} + {1 
      \over n^2} - {9 \over 2} \right\}
      \label{eq:Cn}
 \ee
where $C_F \equiv (N_c^2 - 1) / (2 N_c) \to 4 / 3$ and $S_k(n) \equiv \sum_{j=1}^n 1 /j^k$. For large $n$ (corresponding to the large-$x$ region) the coefficient $C_n^{(NLO)}$ is logarithmically divergent; indeed, since $S_1(n) = \gamma_E + \mbox{log}(n) + O(1/n)$ and $S_2(n) = \pi^2/6 + O(1/n)$ (with $\gamma_E$ being the Euler-Mascheroni constant), one gets
 \be
      C_n^{(NLO)} & = & C_{DIS}^{(NLO)} + C_{n, LOG}^{(NLO)} + O(1/n) ~ ,
      \nonumber \\
      C_{DIS}^{(NLO)} & = & C_F \left[ \gamma_E^2 + {3 \over 2} \gamma_E - 
      {9 \over 2} - {\pi^2 \over 6} \right] ~ , \nonumber \\
      C_{n, LOG}^{(NLO)} & = & C_F ~ \mbox{ln}(n) \left[ \mbox{ln}(n) + 
      2\gamma_E + {3 \over 2} \right]
      \label{eq:Cn_exp}
 \ee

\indent The physical origin of the logarithm and double logarithm terms in Eq. (\ref{eq:Cn}) is the mismatch among the singularities generated in the quark coefficient function by the virtual gluon loops and the real gluon emissions, the latter being suppressed as the elastic peak corresponding to the threshold $x = 1$ is approached. In other words at large $x$ the relevant scale is no more $Q^2$, but $Q^2 (1 - x)$ \cite{Brodsky} and the usual Altarelli-Parisi evolution equation should be accordingly modified \cite{Amati}. The presence of the above-mentioned divergent terms at large $n$ would spoil the perturbative nature of the $NLO$ approximation (\ref{eq:NLO}) (as well as of any fixed-order calculation) and therefore the effects of soft gluon emissions should be considered at all orders in the strong coupling constant $\alpha_s$. To this end one can take advantage of resummation techniques, which show that in moment space soft gluon effects exponenziate \cite{Sterman,Catani_1,Catani_2}. Thus, the moments of the leading twist including the $SGR$ effects, acquire the following form (cf. \cite{Vogt,Sterman00}):
 \be
       \mu_n^{(SGR)}(Q^2) = A_n^{(SGR)} \left[ {\alpha_s(Q^2) \over 
       \alpha_s(Q_0^2)} \right]^{\gamma_n^{NS}} \left\{ \left[1 + 
       {\alpha_s(Q^2) \over 2 \pi} C_{DIS}^{(NLO)} \right]  e^{G_n(Q^2)} + 
       {\alpha_s(Q^2) \over 4 \pi} R_n^{NS} \right\}
      \label{eq:SGR}
 \ee
where $R_n^{NS} = 2 [C_n^{(NLO)} - C_{DIS}^{(NLO)} - C_{n, LOG}^{(NLO)}] + \gamma_n^{(1, NS)} - \beta_1 \gamma_n^{NS} / \beta_0$. The function $G_n(Q^2)$ is the key quantity of the soft gluon resummation and reads as (cf. \cite{Catani_2,Vogt})
 \be
       G_n(Q^2) = \int_0^1 dz {z^{n-1} - 1 \over 1 - z} \left\{ {1 \over 2} 
       B[\alpha_s(Q^2(1 - z))] + \int_{Q^2}^{Q^2(1 - z)} {dq^2 \over q^2} 
      A[\alpha_s(q^2)] \right\}
      \label{eq:Gn_def}
 \ee
where $A[\alpha_s] = C_F \alpha_s / \pi + C_F K (\alpha_s / \pi)^2 / 2$, $B[\alpha_s] = - 3 C_F \alpha_s / 2\pi$ with $K = C_A (67/18 - \pi^2 / 6) - 10 T_R N_f / 9$, $C_A = N_c \to 3$ and $T_R = 1/2$. Explicitly one has
 \be
       G_n(Q^2) = \mbox{ln}(n) G_1(\lambda_n) + G_2(\lambda_n) +O[\alpha_s^k 
       \mbox{ln}^{k-1}(n)]
       \label{eq:Gn}
 \ee
where $\lambda_n \equiv \beta_0 \alpha_s(Q^2) \mbox{ln}(n) / 4\pi$ and
 \be
       G_1(\lambda) & = & C_F {4 \over \beta_0 \lambda} \left[ \lambda + (1 
       -  \lambda) \mbox{ln}(1 - \lambda) \right] ~ , \nonumber \\
       G_2(\lambda) & = & - C_F {4 \gamma_E + 3 \over \beta_0} \mbox{ln}(1 
       - \lambda) - C_F {8 K \over \beta_0^2} \left[ \lambda + \mbox{ln}(1 
       - \lambda) \right] + \nonumber \\
      & & C_F {4  \beta_1 \over \beta_0^3} \left[ \lambda + \mbox{ln}(1 - 
      \lambda) + {1 \over 2} \mbox{ln}^2(1 - \lambda) \right]
     \label{eq:G1G2}
 \ee
It is straightforward to check that in the limit $\lambda_n << 1$ one has $G_n(Q^2) \to \alpha_s(Q^2) C_{n, LOG}^{(NLO)} / 2\pi$, so that at $NLO$ Eq. (\ref{eq:SGR}) reduces to Eq. (\ref{eq:NLO}).

\indent In \cite{Ricco} the Nachtmann moments $M_n(Q^2)$ have been analyzed adopting the $NLO$ approximation (\ref{eq:NLO}) and using the value $\alpha_s(M_Z^2) = 0.113$ derived in  \cite{Virchaux}. However, both a recent re-analysis of the combined $BCDMS$ and $SLAC$ data \cite{Alekhin} and the updated $PDG$ summary \cite{PDG} of various determinations of $\alpha_s(M_Z^2)$ clearly indicates a larger value for $\alpha_s(M_Z^2)$. Therefore, in order to investigate not only the impact of high-order perturbative corrections, but also the relevance of the value of $\alpha_s(M_Z^2)$ we will consider in our analyses two cases: $\alpha_s(M_Z^2) = 0.113$ and $\alpha_s(M_Z^2) = 0.118$.

 \indent The results obtained for the ratio of the quark coefficient function calculated within the $SGR$ and at $NLO$, namely
 \be
       r_n(Q^2) = {\left( 1 + {\alpha_s(Q^2) \over 2\pi} C_{DIS}^{(NLO)} 
      \right) e^{G_n(Q^2)} + {\alpha_s(Q^2) \over 2\pi} \left( C_n^{(NLO)} - 
      C_{DIS}^{(NLO)} - C_{n, LOG}^{(NLO)} \right) \over 1 + {\alpha_s(Q^2) 
      \over 2\pi} C_n^{(NLO)}} ~ ,
     \label{eq:ratio}
 \ee
are reported in Fig. 1. It can be seen that soft gluon effects are quite small for the second and the fourth moments, while for $n \geq 6$ they increase significantly as $n$ increases and exhibit a remarkably sensitivity to the value of $\alpha_s(M_Z^2)$, particularly for $Q^2 \sim$ few $(GeV/c)^2$.

\indent Using the twist expansion (\ref{eq:twists}), the Nachtmann moments $M_n(Q^2)$ of Ref. \cite{Ricco} have been analyzed adopting the $SGR$ expression (\ref{eq:SGR}) for the twist-2 moments and the values $\alpha_s(M_Z^2) = 0.118$ and $Q_0^2 = 10 ~ (GeV/c)^2$. Our results, including the uncertainties for each twist term generated by the fitting procedure\footnote{In our analyses the twist-2 parameter $A_n^{(SGR)}$ [or $A_n^{(NLO)}$ in case of NLO analyses] as well as the higher twist parameters $a_n^{(4)}, \gamma_n^{(4)}, a_n^{(6)}, \gamma_n^{(6)}$ are simultaneously determined from a $\chi^2$-minimization procedure (cf. \cite{Ricco}). We have checked that within the errors generated by the fitting procedure the same value for the leading twist parameter  $A_n^{(SGR)}$ [$A_n^{(NLO)}$] is obtained by limiting the range of our analysis only at large $Q^2$ [namely $Q^2 \gsim n ~ (GeV/c)^2$] without including higher twist terms in Eq. (\ref{eq:twists}).}, are reported in Fig. 2 for $n \geq 4$. It can be seen that: ~ i) the extracted twist-2 term yields an important contribution in the whole $Q^2$-range of the present analysis, and ~ ii) the $Q^2$-behaviour of the data leaves room for a higher twist contribution positive at large $Q^2$ and negative at $Q^2 \sim 1 \div 2 ~ (GeV/c)^2$; such a change of sign requires in Eq. (\ref{eq:twists}) a twist-6 term with a sign opposite to that of the twist-4. As already noted in \cite{Ricco}, such opposite signs make our total higher twist contribution smaller than its individual terms.

\indent We have then carried out twist analyses of the Nachtmann moments $M_n(Q^2)$ adopting the $NLO$ approximation (\ref{eq:NLO})  and using also the value $\alpha_s(M_Z^2) = 0.113$. The comparison of the results obtained in this way, is shown in Figs. 3-4 for the leading twist $\mu_n(Q^2)$ and for the total higher twist contribution $\mbox{HT}_n(Q^2)$, defined as (cf. Eq. (\ref{eq:twists}))
 \be
       \mbox{HT}_n(Q^2) = {Q_0^2 \over Q^2} ~ a_n^{(4)} \left[ 
       {\alpha_s(Q^2) \over \alpha_s(Q_0^2)} \right]^{\gamma_n^{(4)}} + 
       {Q_0^4 \over Q^4} ~ a_n^{(6)} \left[ {\alpha_s(Q^2) \over 
       \alpha_s(Q_0^2)} \right]^{\gamma_n^{(6)}}
       \label{eq:HT} ~ .
 \ee
From Fig. 3 it can clearly be seen that both soft gluon effects and the value of $\alpha_s(M_Z^2)$ can change drastically the evolution of the leading twist at $Q^2 \sim$ few $(GeV/c)^2$ \footnote{The impact of the results shown in Fig. 3 on the large-$x$ behaviour of $PDF$ will be presented elsewhere.}. Correspondingly (see Fig. 4), the role of the higher twists is significantly reduced both by the effects of the soft gluon resummation and by the value adopted for $\alpha_s(M_Z^2)$. On the one hand, within the $NLO$ approximation, the extracted higher twists for $n \geq 6$ can exceed $50 \%$ of the leading twist at $Q^2 \sim$ few $(GeV/c)^2$ both for $\alpha_s(M_Z^2) = 0.113$ and $\alpha_s(M_Z^2) = 0.118$. On the other hand, within the $SGR$ approach, the total higher twist contribution can be still very significant at $Q^2 \sim$ few $(GeV/c)^2$ if the value $\alpha_s(M_Z^2) = 0.113$ is adopted, whereas it does not exceed $\sim 30 \%$ of the leading twist for $Q^2 \gsim 2 ~ (GeV/c)^2$, when the updated $PDG$ value $\alpha_s(M_Z^2) = 0.118$ is considered. Therefore, we can state that the $NLO$ approximation for the leading twist lead to a strong overestimate of the extracted higher twists at large $x$.

\indent Before closing, we want to comment briefly on Eq. (\ref{eq:HT}) as well as on various models used in the literature to describe higher twists at large $x$ (see Refs. \cite{Bodek}-\cite{Alekhin} and \cite{Virchaux}).  As for the latter ones, an interesting way to describe the $x$-shape of higher twists is the renormalon approach \cite{renormalons}, in which both $1 / Q^2$ and $1 / Q^4$ power corrections appear as the result of an ambiguity in the summation of a specific class of high-order radiative corrections. However, the main drawback of the renormalon approach is the fact that the overall strength of the renormalon contribution cannot be predicted theoretically and should be fixed phenomenologically (see Refs. \cite{Bodek,renormalons,Ricco}). We can say that renormalons parameterize our ignorance about high-order perturbative effects, whereas the approach employed in this work (the soft gluon resummation) provides a way to calculate the contribution of a broad class of high-order radiative corrections without introducing any {\em additional} unknown parameter. 

\indent At present, a rigorous $QCD$-based treatment of higher twist operators, including their mixing under the renormalization-group equations, is not completely available. Therefore, one has to resort to phenomenological ans\"atze, like those used in \cite{Kataev,Alekhin,Virchaux} and in the present work, i.e. Eq. (\ref{eq:HT}) (cf. also \cite{Ji,Ricco}). We want to point out that the ans\"atze used in  \cite{Kataev,Alekhin,Virchaux}, which by the way are limited only to a twist-4 term, correspond to special cases of Eq. (\ref{eq:HT}). As a matter of fact, the perturbative evolution of the twist-4 is neglected in \cite{Kataev,Alekhin}, while it is assumed to be the same as the one of the leading twist in \cite{Virchaux}; this corresponds in Eq. (\ref{eq:HT}) to assume $\gamma_n^{(4)} = 0$ and $\gamma_n^{(4)} = \gamma_n^{NS}$, respectively. On the contrary, in \cite{Ricco} as well as in the present work we have found non-vanishing twist-4 (and twist-6) effective anomalous dimensions, which differ significantly from the values of the twist-2 anomalous dimensions. Nevertheless, Eq. (\ref{eq:HT}) is still a crude approximation, because, generally speaking, more than one anomalous dimensions are expected to occur both for the twist-4 and the twist-6 terms. It is therefore clear that any phenomenological ans\"atz may introduce a model dependence in the final result, and in this respect it becomes important to try to estimate the theoretical uncertainties associated to the different higher-twist models as well as to the different perturbative approximations. Such a goal is beyond the aim of the present work and will be the subject of future investigations.

\indent In conclusion, we have analyzed the existing data on the (unpolarised) transverse structure function of the proton for large values of the Bjorken variable $x$ by performing power correction analyses of the Nachtmann moments of Ref. \cite{Ricco} for $1 \lsim Q^2(GeV/c)^2 \lsim 20$. The results obtained adopting the $NLO$ approximation and those including the effects of soft gluon resummation have been compared. We have shown that: ~ i) higher twist extractions based on the $NLO$ approximation for the leading twist are not reliable, and ~ ii) the effects of the soft gluon resummation as well as the updated $PDG$ value of $\alpha_s(M_Z^2)$ should be taken into account at large $x$.

\section*{Acknowledgments}

The author gratefully acknowledges S.J. Brodsky for encouraging his work and S. Forte for many fruitful comments and discussions.

\newpage

\begin{figure}[htb]

\centerline{\epsfxsize=16cm \epsfig{file=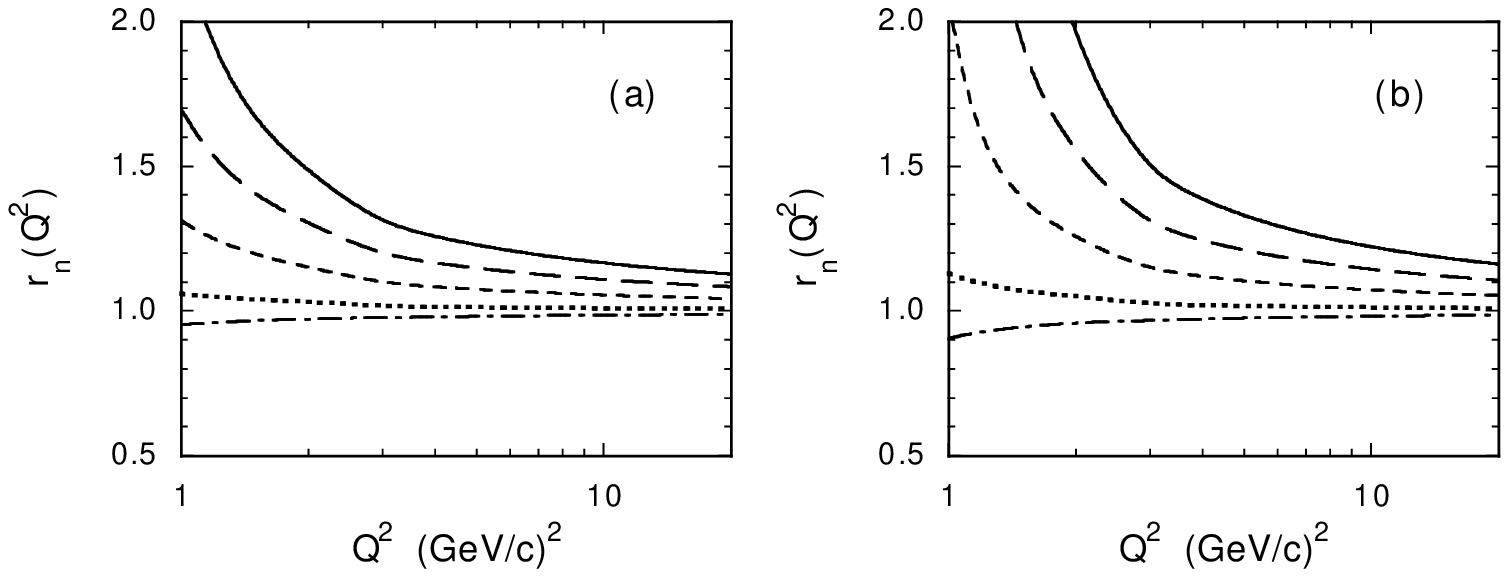}}

\rightline{} \vspace{2cm}

\indent {\bf Figure 1}. Ratio $r_n(Q^2)$ of the quark coefficient functions evaluated within the $SGR$ approach and at $NLO$ (see Eq. (\ref{eq:ratio})) versus $Q^2$, adopting the values $\alpha_s(M_Z^2) = 0.113$ (a) and $\alpha_s(M_Z^2) = 0.118$ (b). Dot-dashed, dotted, short-dashed, long-dashed and solid lines correspond to $n = 2, 4, 6, 8$ and $10$, respectively.

\end{figure}

\newpage

\begin{figure}[htb]

\centerline{\epsfxsize=16cm \epsfig{file=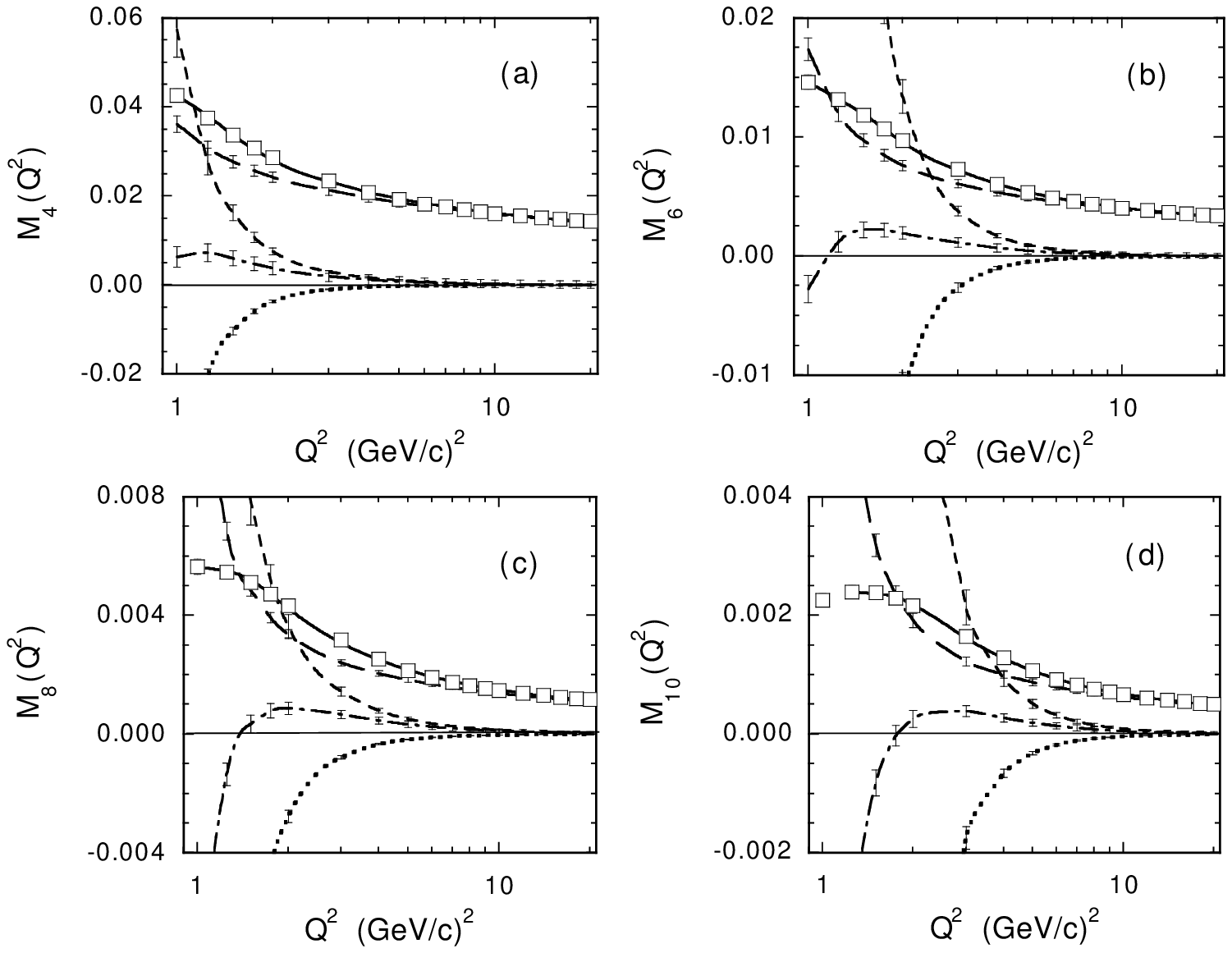}}

\rightline{} \vspace{2cm}

\indent {\bf Figure 2}. Nachtmann moments $M_n(Q^2)$ of the (unpolarised) transverse structure function of the proton versus $Q^2$ for $n = 4$ (a), $n = 6$ (b), $n = 8$ (c) and $n = 10$ (d). The open squares are the {\em "experimental"} moments as determined in \cite{Ricco}, and the solid lines are the results of the twist analysis based on Eq. (\ref{eq:twists}) adopting the $SGR$ expression (\ref{eq:SGR}) for the leading twist. Long-dashed, short-dashed and dotted lines represent the twist-2, twist-4 and twist-6 terms separately, while the dot-dashed lines represent the total higher twist contribution. The error bars on the different twist contributions correspond to the uncertainties generated by the fitting procedure. The value $\alpha_s(M_Z^2) = 0.118$ is adopted. 

\end{figure}

\newpage

\begin{figure}[htb]

\centerline{\epsfxsize=16cm \epsfig{file=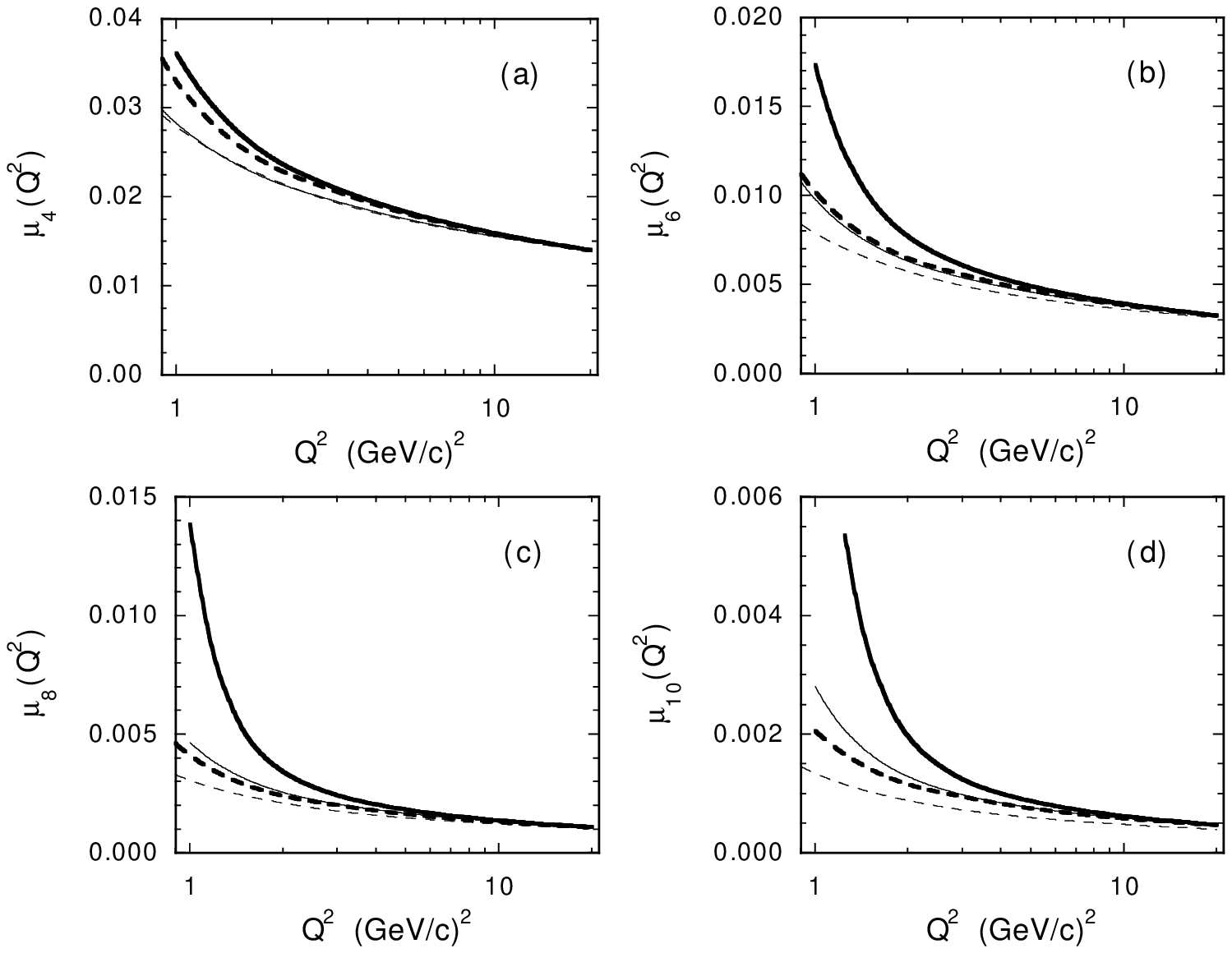}}

\rightline{} \vspace{2cm}

\indent {\bf Figure 3.} The leading twist moments $\mu_n(Q^2)$ resulting from the twist analyses of the Nachtmann moments of Ref. \cite{Ricco}, based on the $SGR$ expression (\ref{eq:SGR}) [solid lines] and on the $NLO$ approximation (\ref{eq:NLO}) [dashed lines]. Thin and thick lines correspond to the cases $\alpha_s(M_Z^2) = 0.113$ and $\alpha_s(M_Z^2) = 0.118$, respectively. The order of the moment is $n = 4$ (a), $n = 6$ (b), $n = 8$ (c) and $n = 10$ (d).

\end{figure}

\newpage

\begin{figure}[htb]

\centerline{\epsfxsize=16cm \epsfig{file=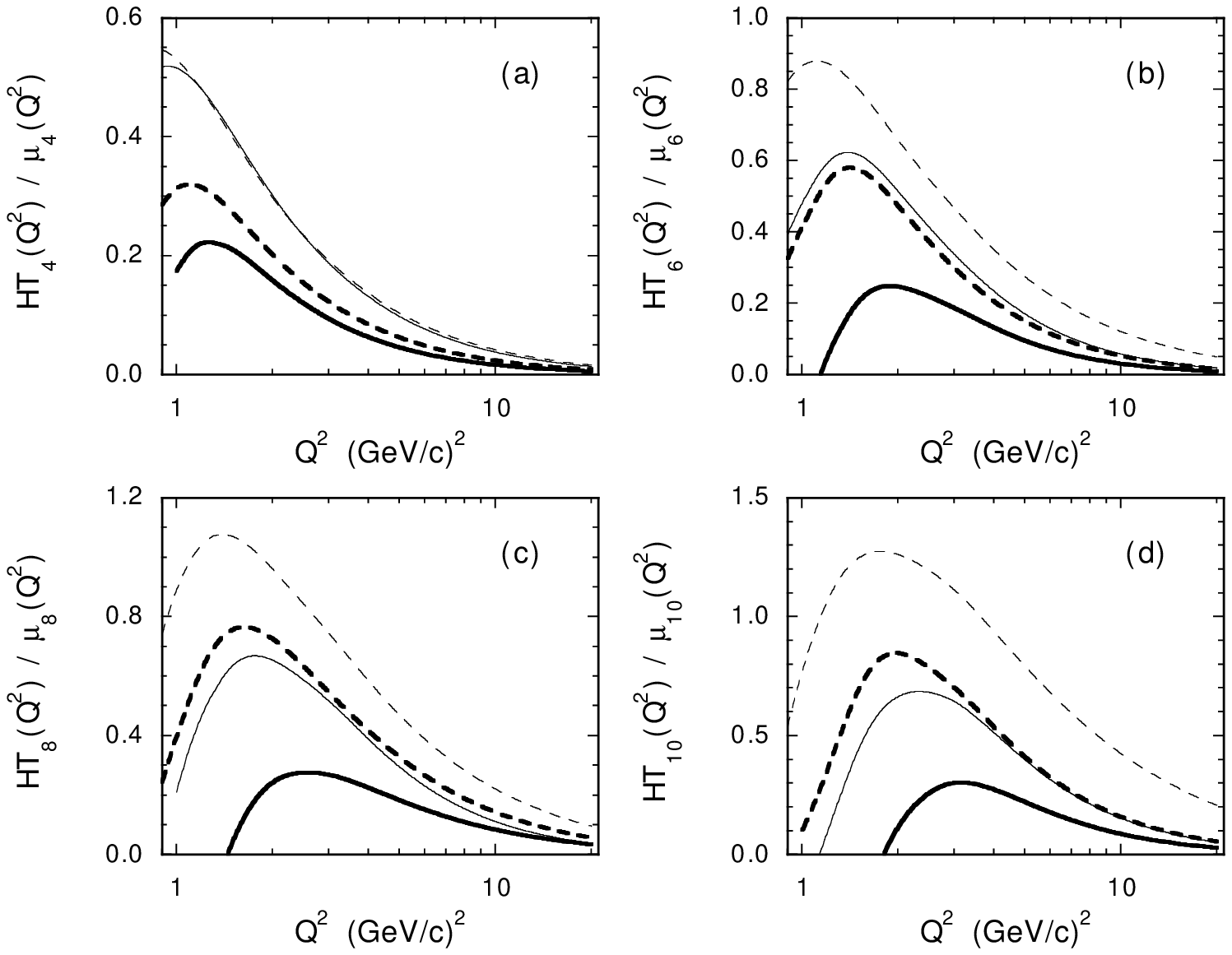}}

\rightline{} \vspace{2cm}

\indent {\bf Figure 4.} The same as in Fig. 3, but for the ratio of the total higher twist contribution $\mbox{HT}_n(Q^2)$ (Eq. (\ref{eq:HT})) to the leading twist term $\mu_n(Q^2)$.

\end{figure}

\end{document}